\documentclass[prl,aps,twocolumn]{revtex4}
\usepackage{epsf,amsmath,color,graphicx,subeqn}
\newcommand {\be}{\begin{equation}}
\newcommand {\ee}{\end{equation}}
\newcommand {\bea}{\begin{eqnarray}}
\newcommand {\eea}{\end{eqnarray}}
\newcommand {\nn} {\nonumber}

\begin{document}

\title{Evaporative Cooling of a Photon Fluid to Quantum Degeneracy}
\author{B. T. Seaman and M. J. Holland}
\affiliation{JILA, National Institute of Standards and Technology and
Department of Physics, \\University of Colorado, Boulder CO
80309-0440, USA}
\date{\today}

\begin{abstract}
  We demonstrate that the process of evaporative cooling, as
  associated with the cooling of atomic gases, can also be employed to
  condense a system of photons giving rise to coherent
  properties of the light. The system we study consists of photons in
  a high-quality Fabry-Perot cavity with photon interactions mediated
  by a nonlinear atomic medium. We predict a macroscopic occupation of the lowest energy mode and evaluate the conditions for realizing a narrow spectral width indicative of a long coherence time for the field.  \pacs{}
\end{abstract}

\maketitle

The phenomenon of coherence has played a crucial role in many areas of
physics. An
extraordinarily long coherence time is the fundamental property that
distinguishes laser light from ordinary light~\cite{Schawlow1958}. The extensive coherence
properties of matter have recently been investigated in trapped
ultracold atoms by studying de Broglie matter waves in
Bose-Einstein condensates~\cite{Anderson1995,Davis1995,Bradley1995}.
Coherence is a ubiquitous phenomenon: many other systems, such as a
collection of polaritons~\cite{Balili2007}, also show similar coherent
attributes.  

In the case of atoms, coherence presents itself when the atomic gas has been condensed to quantum degeneracy.  Condensation is often achieved using the
technique of evaporative cooling in which the highest energy atoms
are forced to escape the trap, and those that remain rethermalize, thereby reducing the temperature.
In a nonequilibrium
system, continuous evaporative cooling may generate a coherent atom
laser~\cite{Holland1996}. In this article, we demonstrate how
evaporative cooling and Bose-stimulated emission can be used to
condense a photon fluid into a quantum degenerate superfluid.

In a
general sense, a photon fluid is a collection of spatially localized
photons which, through their interactions via a nonlinear medium,
exhibit hydrodynamic or similar fluidic behavior. 
A condensed photon fluid should possess superfluid properties such
as coherence, phase rigidity, quantized vortices, a critical Landau velocity,
and the Bogoliubov dispersion~\cite{Chiao1999}. This is distinct
from a normal laser, where the effective photon interactions are so weak that they are unimportant.  Another significant difference with usual lasers is
that a population inversion of an internal
atomic state is not necessary to generate the coherent light. Instead the phase-space compression to produce a
macroscopically occupied mode relies on an inversion of
photon cavity mode populations with respect to the Planck distribution.  Since the relevant
modes of the system are determined by the structural properties of the
cavity and not atomic quantities, the frequency of the coherent field could be highly tunable.

Two high-reflectivity mirrors placed close together form a Fabry-Perot
cavity and support many photon modes. The closeness of the mirrors
creates a cavity mode volume which has a high aspect ratio, with large
energy gaps between modes which have adjacent longitudinal quantum
numbers and small energy gaps between modes which have adjacent
transverse quantum numbers. We utilize this to pump the cavity
such that only one longitudinal quantum number is relevant, and
the modes considered are distinguished only by their
transverse degrees of freedom. Photons in vacuum interact
weakly, therefore, it is necessary to incorporate a nonlinear medium, such as an
atomic Rydberg gas or nonlinear crystal, into the cavity to
allow for atom-mediated photon-photon interactions.
Since the longitudinal degree of freedom is frozen out due to a small
cavity length, an intriguing link exists between this system and condensation in
two dimensions with the well-known physics of the
Berezinskii-Kosterlitz-Thouless transition, which has recently been
investigated in atomic gases~\cite{Hadzibabic2006,Schweikhard2007}.

\begin{figure}
\begin{center}
  \epsfxsize=7cm \leavevmode \epsfbox{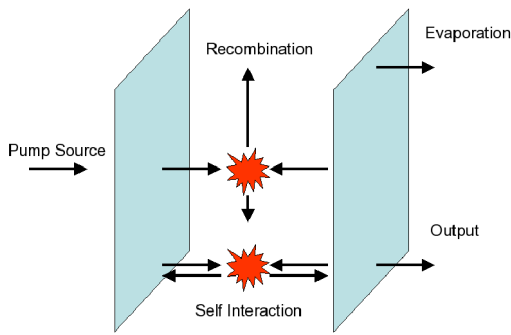}
\caption{(Color online) 
The five primary processes that occur in the Fabry-Perot cavity are
presented.  Photons enter the system from a pumping source.  While in
the cavity, the photons recombine and self-interact.  Finally, the
photons can leave the cavity through evaporation or output coupling.
\label{fig:Cavity}}
\end{center}
\end{figure}

The photon fluid in a Fabry-Perot cavity is governed by the
Hamiltonian,
\begin{equation}
  \hat{H}=\sum_i\hbar\omega_i\hat{a}^\dag_i\hat{a}_i
  +\sum_{ijkl}\hbar\Gamma^C_{ijkl}\hat{a}^\dag_i\hat{a}^\dag_j\hat{a}_k\hat{a}_l\, ,
\label{eq:h0}
\end{equation}
where $\hat{a}_i$ is the annihilation operator for a photon in mode
$i$ with energy $\hbar\omega_i$, and $\Gamma^C_{ijkl}$ is the
scattering or collision ($C$) rate for photons from modes $k$ and $l$
into modes $i$ and $j$. There are two kinds of interactions, as
illustrated in Fig.~\ref{fig:Cavity}.
Terms which do not change
populations give rise to a self-interaction energy and leave the
photons in the same mode they entered in, e.g.
$\hat{a}^\dag_0\hat{a}^\dag_1\hat{a}_1\hat{a}_0$. Population transfer
is due to recombination terms, such as
$\hat{a}^\dag_0\hat{a}^\dag_2\hat{a}_1\hat{a}_1$.

Photons can also enter and leave the cavity through irreversible
pumping and decay, which is not described by the Hamiltonian 
in Eq.~(\ref{eq:h0}). Pumping and decay can be incorporated
into the system dynamics through the quantum master-equation formalism,
\begin{equation}
 \frac{\partial\hat{\rho}}{\partial
  t}=-\frac{i}{\hbar}[\hat{H},\hat{\rho}]
+\sum_i\Gamma^P_i\hat{L}[\hat{a}_i^\dag]
+\sum_i\Gamma^D_i\hat{L}[\hat{a}_i]\, ,
\label{eq:denmat}
\end{equation}
where $\hat{\rho}$ is the density matrix operator of the system,
$\Gamma^P_i$ is the pumping rate ($P$) and $\Gamma^D_i$ is,
the decay rate ($D$) for mode $i$, and the Lindbladt superoperator is
 \begin{equation}
   \hat{L}[\hat{O}]\equiv2\hat{O}\hat{\rho}\hat{O}^\dag-\hat{O}^\dag\hat{O}
   \hat{\rho}-\hat{\rho}\hat{O}^\dag\hat{O}\, .
\end{equation}
The relative pumping rates as a function of the frequency of the system modes are
dependent on the pumping source, which, for simplicity, we take here to be constant across all relevant modes, $\Gamma^P_i\rightarrow\Gamma^P$.

In contrast, evaporative cooling requires a
decay rate from the cavity which is strongly dependent on the energy.
A mirror with a frequency dependent reflectivity, which allows
modes to decay at different rates, could be utilized.  A reflectivity which drops for large frequencies can create
conditions where high energy photons leave the cavity faster than low
energy photons. Since a high energy photon carries away more than the
average photon energy, subsequent rethermalization through photon
interactions should reduce the system temperature, in much the same way as
in the evaporative cooling of atoms.

We implement energy dependent reflectivity by dividing the decay rates into two classes.
High energy photons ($H$) are described by a decay rate of $\Gamma^H$
and low energy photons ($L$) by a decay rate of $\Gamma^L$, with
$\Gamma^H\gg\Gamma^L$. The large decay rate of the high energy modes
means they can be eliminated in the following way.
Consider an interaction in which photons from modes $i$ and $j$
create a photon of low energy in mode $k$ and a photon of high energy
in mode $l$. The rapid decay of mode $l$ causes this process to be
effectively irreversible. This is described most simply by extracting
such processes which involve a high energy mode from the reversible
$\hat H$ term in Eq.~\ref{eq:denmat}, and adding them back to the
density matrix evolution as an irreversible term of superoperator form,
\begin{equation}
  \Gamma^E \hat{L}[\hat{a}^\dag_k\hat{a}_i\hat{a}_j]\, ,
\end{equation}
where $\Gamma^E\equiv(\Gamma^C)^2/\Gamma^H$ is the evaporation rate
($E$), taken to be constant. The system Hilbert
space is now reduced and the high energy modes no longer appear in the
theory explicitly.

Even with these simplifications, the problem is still intractable in
general. The dimensionality of the Hilbert space grows exponentially
with the number of photons in the lower energy modes. Most of this complexity is uninteresting from the point of
view of evaporative cooling of photons since it arises from all 
possible pathways to redistribute and entangle photons amongst the lowest energy
modes. Consequently, we divide the system into what we will refer to as plaquettes, each
plaquette consisting of a pair of modes $i$ and $j$, which
may interact to produce a photon in the lowest energy mode $k=0$ and a
photon in a high energy mode $l$ which is lost through the mirrors
rapidly. Adding the contributions from $D$ distinct plaquettes as an
incoherent sum of pumping rates into the ground state, 
the density matrix equation of motion in the interaction picture is,
\bea
\label{eq:DensityMatrixEquations}
\frac{\partial\hat{\rho}}{\partial
  t}&=&
+\Gamma^P\sum_{i}\hat{L}[\hat{a}^\dag_{i}]+\Gamma^L\sum_{i}\hat{L}[\hat{a}_{i}]\\
&&
-i\Gamma^C\sum_i[
\hat{a}^\dag_0 
  \hat{a}^\dag_{i}\hat{a}_{i}\hat{a}_0
,\hat{\rho}]
+\Gamma^E\sum_{\langle ij\rangle}\hat{L}[\hat{a}^\dag_0\hat{a}_i\hat{a}_j]\nn\, ,
\eea
where the sum over $\langle i,j\rangle$ implies modes $i$ and $j$ are from the same plaquette.

Amongst the populations, the evolution time scales are
not all equivalent. If only one or no atoms are in a plaquette, a
slow evolution takes place on time scales governed by $\Gamma^P$ and $\Gamma^L$. If two atoms are in a plaquette, however, and providing $\Gamma^E\gg\Gamma^P,\Gamma^L$, they will rapidly collide and form a ground state photon, and a high energy photon which will evaporate away.  Consistent with this is that we may neglect the possibility of three or more photons in the plaquette.

The basis states needed then describe the number of photons in each plaquette and the number of photons in the lowest energy mode.  The precise occupation of the modes is not important, only how many plaquettes have a given number of photons.  For instance, an arbitrary basis state $|\Psi\rangle$ can be uniquely identified with a reduced computational state described by,
\be
|\Psi\rangle\to|n,abcd\rangle\equiv|n,(00)^a(01)^b(11)^c(02)^d\rangle\, ,
\ee
where
$n$ is the number of ground state photons, $a$ is the number of
plaquettes with no photons, $b$ is the number of plaquettes with one
photon, and $c$ and $d$ are the number of plaquettes with two photons in
different modes or the same mode, respectively.  

The photon number distribution is an important quantity to characterize the photon state since a coherent
state has a Poissonian distribution while a thermal state has an
exponentially decaying distribution.
The populations of
all states $|\Psi\rangle$ that can be expressed as $|n,abcd\rangle$ are given by
\be
\label{eq:Pops}
P_{n,abcd}\equiv\langle \Psi|\rho|\Psi\rangle\, . \ee
We can now take advantage of the natural separation of
time scales by adiabatically eliminating the fast processes.
This is implemented by solving for the fast variables in steady state and substituting back into the equations for the slow variables.
The populations of states with a plaquette with two photons are then approximately given by their
steady state values, \bea
P_{n,ab10}&=&\frac{2}{n+1}\frac{\Gamma^P}{\Gamma^E}P_{n,a(b+1)00}\, ,\\
P_{n,ab01}&=&\frac{1}{n+1}\frac{\Gamma^P}{\Gamma^E}P_{n,a(b+1)00}\, ,
\eea as determined by Eq.~\ref{eq:DensityMatrixEquations} and
Eq.~\ref{eq:Pops}. The evolution of the remaining
populations is then determined by, \bea \frac{\partial
  P_{n,a}}{\partial t}&=&
-2(n+D-a)\Gamma^LP_{n,a}\\
&&-2(n+1+3D-a))\Gamma^PP_{n,a}\nn\\
&&+2(n+1)\Gamma^LP_{n+1,a}+2n\Gamma^PP_{n-1,a}\nn\\
&&+2(D-a)\Gamma^PP_{n,a+1}+4a\Gamma^LP_{n,a-1}\nn\\
&&+12a\Gamma^PP_{n-1,a-1}\, ,\nn \eea where $P_{n,a}\equiv
P_{n,a(D-a)00}$.

\begin{figure}[t]
\begin{center}
\epsfxsize=7cm \leavevmode \epsfbox{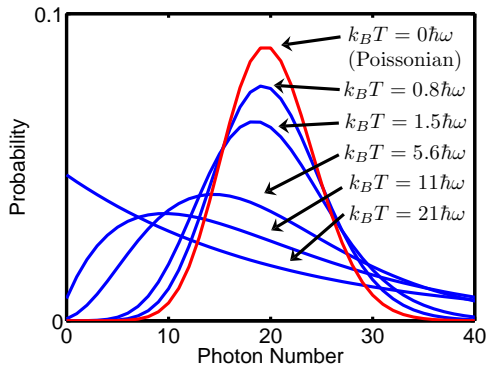}
\caption{
(Color online) Photon number probabilities of the lowest photon mode
of steady states with constant number of low energy photons.  As the
pumping power per mode is decreased and more modes are pumped, denoted
by decreasing temperature, the distributions change from an exponential decay
associated with a thermal system to nearly Poissonian as associated
with a superfluid system.  
\label{fig:DThreeModePhotonNumber}}
\end{center}
\end{figure}

The number distribution of photons in the low energy mode is highly
dependent on the incoherent pumping rate and the number of plaquettes
pumped.  The pumping rate can be designated by an effective temperature, $T$, \be
e^{-\hbar\omega/k_BT}\equiv\frac{\Gamma^P}{\Gamma^L}=\frac{\bar{n}}{\bar{n}+1}\,
, \ee where $\bar{n}$ is the number of photons that would be in each
mode if the modes were isolated and non-interacting and $\omega$ is
the bare mode energy.  This temperature corresponds to the occupancy of the external pumping modes.
In order to make an effective comparison between different temperatures, we hold constant the average number of photons in the lowest energy mode, and vary both the temperature and, correspondingly, the energy width of the pump in order to maintain this condition.
It would be expected that for an intense, narrow-bandwidth pump connected only with the lowest energy mode of the Fabry-Perot cavity, the high temperature and incoherent characteristics of the pump would be imprinted on the photons in that mode.  However, for a weaker but broader pump, in which a wider spectrum of modes are pumped but at a slower rate so that the average photon number remains the same, the lowest energy mode is not as influenced by the temperature of the pumping modes.  The occupation of the lowest energy mode is due mainly to recombination and stimulated emission which should create a coherent system in direct analogy with the evaporative cooling of atoms.

Figure~\ref{fig:DThreeModePhotonNumber} presents the number
distributions of photons in the lowest energy mode during steady state
operation for systems with 20 photons in the low energy mode.
With a large effective temperature, as denoted
in the figure by $k_BT=21\hbar\omega$, the number distribution is essentially
what would be expected for a thermal source.  As the effective
temperature decreases, by decreasing the pumping strength and
increasing the number of pumped plaquettes, the effects of bosonic
amplification dramatically increase.  This leads to a nearly
Poissonian number distribution associated with zero temperature
condensates.  The key to converting the incoherent thermal pump into a coherent system is that the occupation of the lowest energy mode comes from stimulated emission from higher energy modes.

The spectrum of the photons is also an important quantity to examine
as it provides information about the coherence time and spectral width. Since the mode
separation is greater than the collisional interaction energy and the
system is in steady state, the fluctuation spectrum is determined by,
\be S(\nu)=\int_{-\infty}^{\infty}d\tau e^{-i\nu\tau}
\frac{\left\langle \hat{a}^\dag_0(\tau) \hat{a}_0(0) \right\rangle}
{\left\langle \hat{a}^\dag_0(0) \hat{a}_0(0) \right\rangle}\, , \ee
which is the Fourier transform of a two-time correlation function.  The
correlation function can be expressed as, \be \langle
\hat{a}_0^\dag(t)\hat{a}_0(0)\rangle
=\sum_{n,\{i\}}\sqrt{n}C_{n,\{i\}}(t)\, , \ee with, \be
C_{n,\{i\}}(t)\equiv
Tr[\hat{\rho}|n,\{i\}\rangle\langle n-1,\{i\}|(t)\hat{a}_0(0)]\, ,
\ee where $n$ is the number of ground state photons and $\{i\}$
represents the configuration of the remaining modes.  From the
quantum regression theorem, the equation of motion for the quantity
$C_{n,\{i\}}$ is given by, \be \frac{\partial C_{n,\{i\}}}{\partial
  t}=\sum_{m\{j\}}M_{n\{i\}m\{j\}}C_{m,\{j\}}\, , \label{eqn:Ceqn}\ee where the matrix
elements are determined from the off-diagonal density matrix equation
of motion, \be \frac{\partial \rho_{n\{i\},(n-1)\{i\}}}{\partial
  t}=\sum_{m\{j\}}M_{n\{i\}m\{j\}}\rho_{(m-1)\{j\},m\{j\}}\, , \ee
as given by Eq.~\ref{eq:DensityMatrixEquations}.  As this is a reduced set of the full density matrix equations of motion, it is merely necessary to replace these matrix elements into Eq.~(\ref{eqn:Ceqn}) with the initial condition
$C_{n,\{i\}}(0)=\sqrt{n}P_{n,abcd}(0)$, where $|n,\{i\}\rangle$ corresponds to the reduced basis state $|n,abcd\rangle$.

There are two driving forces of decoherence, incoherent transfer
through the mirrors and phase dispersion due to collisions, which both limit the coherence time.  The decoherence due to transfer through the
mirrors can be examined by setting a vanishing collision rate.  Due to
bosonic amplification, the photons are able to support a state with a
spectral width much narrower than that given by the linewidth of the
photon decay.

\begin{figure}[t]
\begin{center}
\epsfxsize=7cm \leavevmode \epsfbox{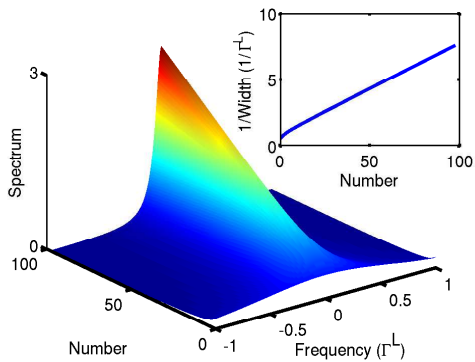}
\caption{
(Color online) The spectrum is presented with vanishing
collision rates.  As the photon number increases, the spectral width narrows.  In the inset, the inverse of the spectral
width is proportional to average photon number as
expected from the Schawlow-Townes linewidth.
\label{fig:DThreeModeSpectrum}}
\end{center}
\end{figure}

Figure~\ref{fig:DThreeModeSpectrum} presents the spectrum for
vanishing collision rates for systems with 20 pumped plaquettes and
different average photon numbers. As the photon number is increased
due to a higher pumping rate, the spectral width decreases due to
Bosonic amplification.  The spectrum can be much narrower than
the width associated with the decay rate through the mirrors.  As shown
in the inset of Fig.~\ref{fig:DThreeModeSpectrum}, the
spectral width, defined as half width at half maximum, is inversely proportional to the average photon number as expected by the
Schawlow-Townes linewidth~\cite{Schawlow1958}.

\begin{figure}[t]
\begin{center}
\epsfxsize=6.5cm \leavevmode \epsfbox{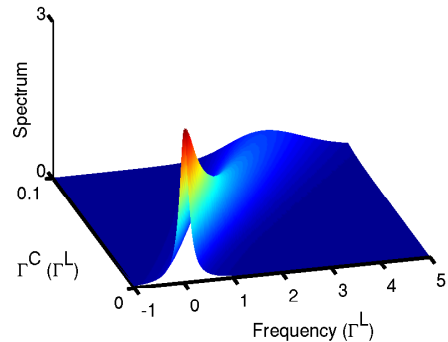}
\caption{
(Color online) The spectrum is
presented for systems with different collision rates.  Larger
collision rates lead to wider spectra.  The spectrum center is also
increased from the bare mode frequency due to mean-field interaction
shifts.
\label{fig:DThreeModeCollisionSpectrum}}
\end{center}
\end{figure}

The effects of non-vanishing collision rates are of course an important aspect of
the system.
Figure~\ref{fig:DThreeModeCollisionSpectrum} presents the fluctuation
spectrum for a system with an average of
$100$ photons as the collision rate is increased. There is a
mean-field interaction shift of the frequency since the energy
of the photons is no longer just the bare mode energy.  The
spectral width rapidly grows as the collision rate becomes larger.
One possible way to remedy this problem is to create a system
that has a small collision
rate for the low energy photons but still possesses a large
recombination rate for the high energy photons.  Such a system can be manufactured by creating a spatially
dependent nonlinear medium.  The low energy photons are more localized
to the center of the cavity than the high energy photons.  Therefore,
if the medium is less dense in the center of the trap than toward the
edges, the cavity will allow for fast recombinations of high energy
photons to promote rapid evaporative cooling but small collision rates
for low energy photons to promote coherent evolution.

The condensation of a photon fluid using evaporative cooling provides
a new mechanism for creating a high-intensity narrow-spectrum coherent
beam of light.  Unlike a laser, the photon fluid, due to the presence
of interactions, is expected to be superfluid and should, for example, exhibit signature properties such as phase rigidity of the order parameter, and the ability to
support quantized vortices.  

This work was supported by the National Science Foundation and the U.S. Department of Energy, Office of Basic Energy Sciences via the Chemical Sciences, Geosciences, and Biosciences Division.

\bibliographystyle{prsty}

\begin{thebibliography}{1}

\bibitem{Schawlow1958}
A.~L. Schawlow and C.~H. Townes, Phys. Rev. {\bf 112},  1940  (1958).

\bibitem{Anderson1995}
M.~H. Anderson {\it et~al.}, Science {\bf 269},  198  (1995).

\bibitem{Davis1995}
K.~B. Davis {\it et~al.}, Phys. Rev. Lett. {\bf 75},  3969  (1995).

\bibitem{Bradley1995}
C.~C. Bradley, C.~A. Scakett, J.~J. Tollett, and R.~G. Hulet, Phys. Rev. Lett.
  {\bf 75},  1687  (1995).

\bibitem{Balili2007}
R. Balili {\it et~al.}, Science {\bf 326},  1007  (2007).

\bibitem{Holland1996}
M. Holland {\it et~al.}, Phys. Rev. A {\bf 54},  R1757  (1996).

\bibitem{Chiao1999}
R.~Y. Chiao and J. Boyce, Phys. Rev. A {\bf 60},  4114  (1999).

\bibitem{Hadzibabic2006}
Z. Hadzibabic {\it et~al.}, Nature {\bf 441},  1118  (2006).

\bibitem{Schweikhard2007}
V. Schweikhard, S. Tung, and E.~A. Cornell, Phys. Rev. Lett. {\bf 99},  030401
  (2007).

\end{thebibliography}

\end{document}